\documentclass[aip,apl,numerical,reprint,superscriptaddress]{revtex4-1}

\usepackage{graphicx}
\usepackage{latexsym}
\usepackage{amsmath}
\usepackage{amssymb}
\usepackage{amsfonts}
\usepackage{color}

\begin{document}
\bibliographystyle{aipsamp}

\title{Magnetic Flux Pumping in Superconducting Loop Containing a Josephson $\psi$ Junction}

\author{S. Mironov}
\affiliation{Institute for Physics of Microstructures, Russian Academy of Sciences, GSP-105, 603950 Nizhny Novgorod, Russia}
\author{H. Meng}
\affiliation{University Bordeaux, LOMA UMR-CNRS 5798, F-33405 Talence Cedex, France}
\affiliation{School of Physics and Telecommunication Engineering, Shaanxi University of Technology, Hanzhong 723001, China}
\affiliation{Shanghai Key Laboratory of High Temperature Superconductors, Shanghai University, Shanghai 200444, China}
\author{A. Buzdin}
\affiliation{University Bordeaux, LOMA UMR-CNRS 5798, F-33405 Talence Cedex, France}
\affiliation{Sechenov First Moscow State Medical University, Moscow, 119991, Russia}

\begin{abstract}
We demonstrate that a Josephson junction with a half-metallic weak link integrated into the superconducting loop enables the pumping of magnetic flux piercing the loop. In such junctions the ground state phase $\psi$ is determined by the mutual orientation of magnetic moments in two ferromagnets surrounding the half-metal. Thus, the precession of magnetic moment in one of two ferromagnets controlled, e.g., by the microwave radiation, results in the accumulation of the phase $\psi$ and subsequent switching between the states with different vorticities. The proposed flux pumping mechanism does not require the application of  voltage or external magnetic field which enables the design of electrically decoupled memory cells in superconducting spintronics.
\end{abstract}

\maketitle

When a superconducting loop is put in the external magnetic field the magnetic flux $\Phi$ through the loop is quantized being equal to the integer number of the flux quanta $\Phi_0$: $\Phi=n\Phi_0$ \cite{Barone}. The interplay between the states with different vorticities $n$ enables the implementation of multiply connected superconducting systems as flux qubits \cite{Orlando_qubit, Makhlin_RMP}, ultra-sensitive magnetic field detectors \cite{Clarke, Fagaly_SQUID},  generators of ac radiation \cite{Solinas} etc. The flux quantization also naturally suggests using superconducting loop as topologically protected memory cell in the devices of the rapid single flux quantum (RSFQ) logics \cite{Likharev_RSFQ}. 

The straightforward way to switch the loop between the states with different vorticity $n$ requires application of external magnetic flux $\Phi_{ext}\sim\Phi_0$. Even for the loops of the radius $R\sim 0.1~{\rm \mu m}$ the corresponding magnetic field should be $\sim 0.1~{\rm T}$ and it grows $\propto R^{-2}$ with the decrease of the loop size which becomes a serious obstacle for the miniaturization of the flux-based devices.  Thus, the discovery of the physical phenomena which enable the controllable switching of the vorticity $n$ without application of external flux is strongly desired. 

One of promising mechanisms allowing the flux-free switching can be realized in the loops containing a Josephson junction with embedded magnetic order and/or broken inversion symmetry \cite{Golubov_RMP, Buzdin_RMP,Linder_rev,Eschrig_rev}. These systems are known to reveal unusual collective dynamics of magnetic and superconducting order parameters.\cite{Konschelle, Volkov1, Volkov2, Cai, Maekawa1, Maekawa2, Hoffman, Bobkova}  In contrast to the conventional Josephson systems, such junctions support the non-zero ground state phase $\varphi_0$ between the superconducting leads. Being incorporated into the loop such $\varphi_0$-junctions play the role of the phase batteries producing the spontaneous electric current  which corresponds to the magnetic flux $\Phi=(\varphi_0/2\pi)\Phi_0$ through the loop \cite{Ustinov,Bauer, Buzdin_2005, Feofanov, Ortlepp}. One can expect, thus, that controlling the ground state phase $\varphi_0$, e.g., by voltage or radiation should enable effective tuning of the magnetic flux and switching the system between the states with different vorticity without application of external magnetic field. 

\begin{figure}[t!]
\includegraphics[width=0.45\textwidth]{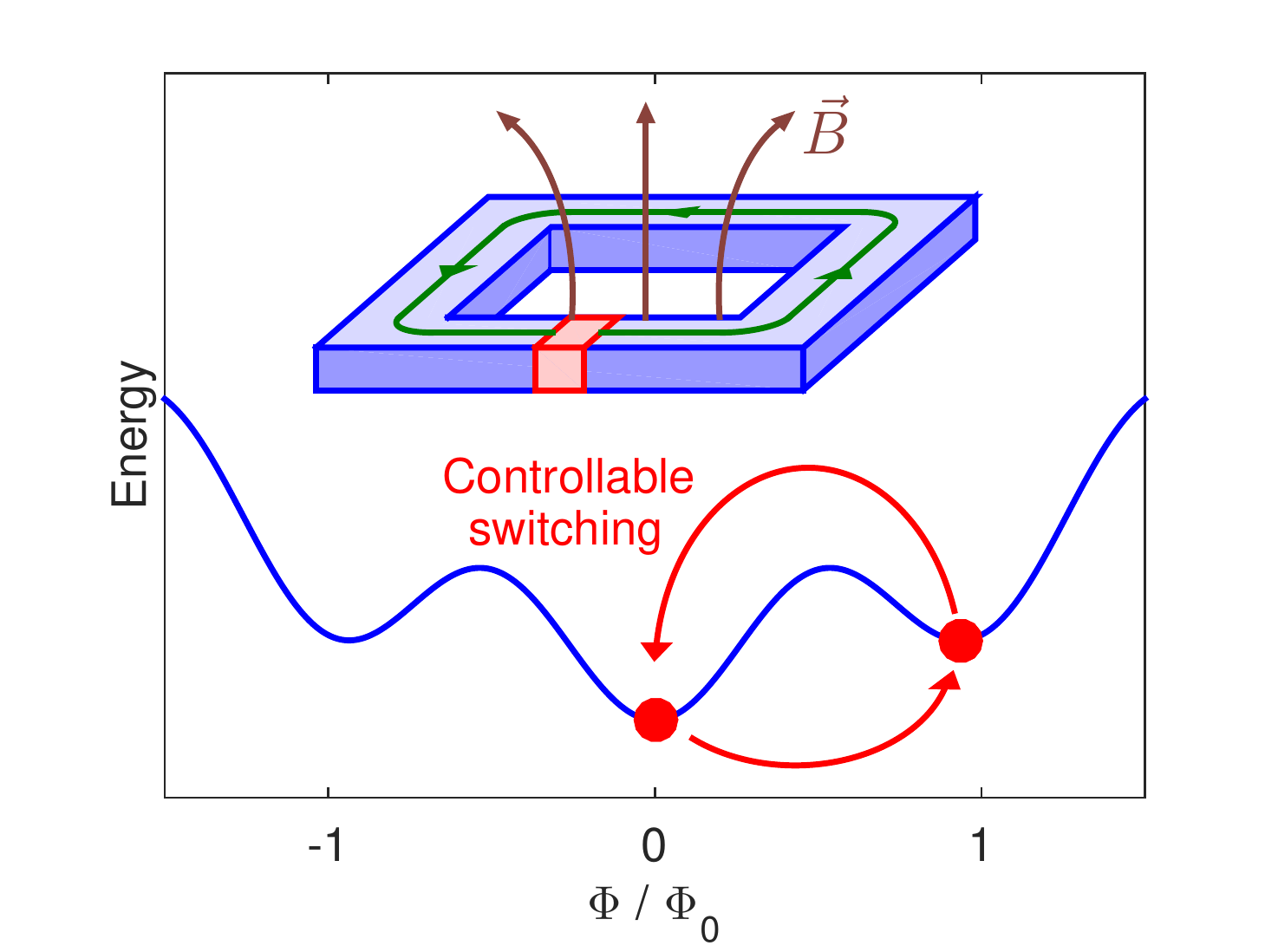}
\caption{Sketch of the superconducting loop containing a Josephson $\psi$ junction which enables the switching between the states with different vorticities by periodic driving of the ground state Josephson phase $\psi$.} \label{Fig1}
\end{figure}

\begin{figure*}[t!]
\includegraphics[width=0.4\textwidth]{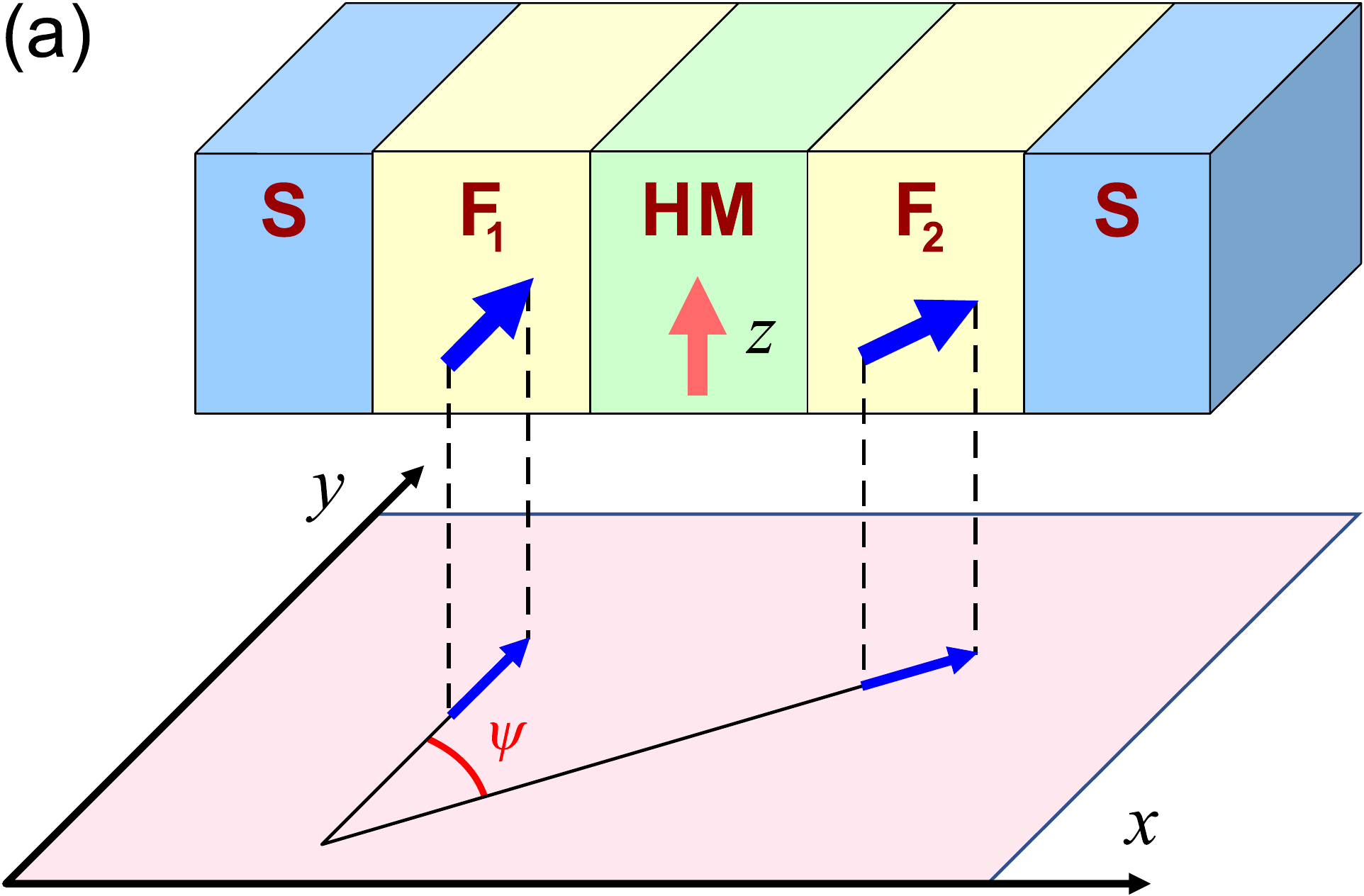}~~~~~~~~~~~~~
\includegraphics[width=0.49\textwidth]{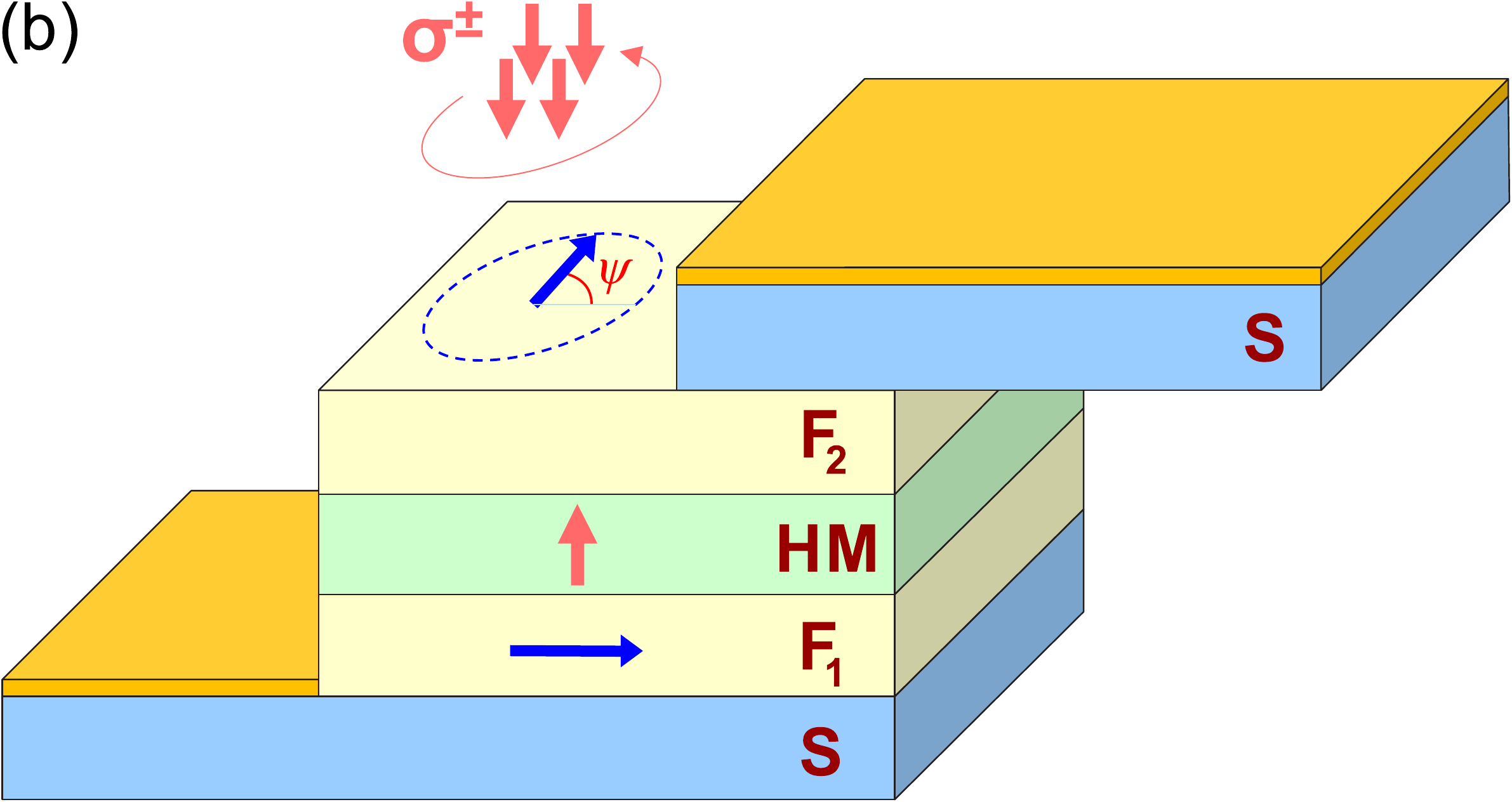}
\caption{(a) Sketch of the Josephson $\psi$ junction where the weak link consists of a half-metallic layer sandwiched between two ferromagnets. The ground state phase $\psi$ coincides with the angle between the projection of magnetic moments in ferromagnets to the plane perpendicular to the spin quantization axis of half-metal. (b) Sketch of possible experimental setup where circularly polarized radiation controls the rotation of magnetic moment in the F$_2$ layer and subsequent variation of the spontaneous Josephson phase $\psi$. The superconducting electrodes are covered by the reflecting material to minimize  heating effects.} \label{Fig2}
\end{figure*}

However, in most existing $\varphi_0$ junctions the phase variation is limited ($\delta\varphi_0<2\pi$) and the corresponding flux $\Phi<\Phi_0$ cannot change the vorticity $n$ of the superconducting state. Indeed, the junctions with a ferromagnetic (F) layer between the superconducting (S) leads support only $\pi$ states with $\varphi_0=\pi$ \cite{Buzdin_pi,Ryazanov_1,Ryazanov_2} which allow the design of the environmentally decoupled qubits \cite{Feofanov} but are not suitable for changing the loop vorticity. The alternating F layer thickness (see, e.g., Ref.~\onlinecite{Goldobin_1} and references therein) as well as the presence of the Abrikosov vortex or the pair of current injectors in one of S leads \cite{Gaber,Ustinov_inj,Goldobin_2} may produce the second order phase transition with the spontaneous phase $\varphi_0$ gradually changing between zero and $\pi$ (so that $\delta\varphi_0<\pi$). The situation looks more promising for the the S/F/S junctions with the broken inversion symmetry where the spin-orbit coupling (SOC)  produces the arbitrary spontaneous phase $\varphi_0\propto\nu\sin\theta$ where $\theta$ is the angle between the exchange field in the F material and the direction of the broken inversion symmetry while the constant $\nu$ characterizes the strength of SOC and the exchange field \cite{Buzdin_Phi, Reynoso, Zazunov, Mironov_Phi, Kouwenhoven_Phi}. However, the parameter $\nu$ is typically small which limits the $\varphi_0$ variations.

In this Letter we unravel the flux-free mechanism of the remote control of the superconducting state vorticity in the loop by the Josephson $\psi$ junction in which the range of the spontaneous phase $\psi$ variation is not limited (see Fig.~\ref{Fig1}). The weak link of the basic $\psi$ junction consists of the layer of half-metal (HM) sandwiched between two ferromagnets (Fig.~\ref{Fig2}a). If the magnetic moments in the three magnetic layers are non-coplanar the current-phase relation of the junction takes the form $I(\varphi)=I_c\sin(\varphi-\psi)$ where the spontaneous phase $\psi$ and the critical current $I_c$ are controlled by the directions of the magnetic moments ${\bf M}_1$ and ${\bf M}_2$ in the F$_1$ and F$_2$ layers. Specifically, for the spin quantization axis in the half-metal directed along the $z$-axis the phase $\psi$ coincides with the angle between the projection of ${\bf M}_1$ and ${\bf M}_2$ to the $xy$-plane while the critical current $I_c\propto\sin\alpha_1\sin\alpha_2$ where $\alpha_1$ ($\alpha_2$) is the angle between ${\bf M}_1$ (${\bf M}_2$) and the $z$-axis.\cite{Braude, Eschrig_JLTP, Eschrig_NatPhys, Grein, Eschrig_NJP, Mironov_HM}. The geometry of the $\psi$ junction favorable for the control of the loop vorticity requires the direction of ${\bf M}_1$ to be fixed along the $y$-axis and the F$_2$ layer to have the easy-plane ($xy$) magnetic anisotropy. In this case radiating the F$_2$ layer with the circularly polarized electromagnetic wave with the magnetic field rotating in the $xy$ plane should cause the precession of ${\bf M}_2$ around the $z$-axis \cite{Bloch} resulting in the unlimited accumulation of the spontaneous phase $\psi$. If the described $\psi$ junction is integrated in the superconducting loop such phase accumulation gives rise to the switching between the states with different vorticities and subsequent pumping of magnetic flux through the loop without actual application of magnetic field.

The described phase accumulation in $\psi$ junctions reminds the Berry phase effect \cite{Xiao_RMP} since the full recovery of the initial magnetic state after the one precession period is accompanied by the $2\pi$ change of the Josephson ground state phase $\psi$. Note that this phenomenon contrasts to the behavior of the $\varphi_0$ junctions where the recovery of the exchange field direction (the angle $\theta$) unavoidably results in the recovery of the spontaneous phase $\varphi_0\propto\sin\theta$.

For the experimental realization of the described radiation-stimulated vorticity switching it is convenient to use the Josephson $\psi$ junctions of the ``overlap'' geometry  (see Fig.~\ref{Fig2}b). This design provides a convenient access to the surface of the F$_2$ layer for the circularly polarized electromagnetic wave. At the same time, the superconducting electrodes should be covered by the suitable reflecting material to prevent parasitic heating effects. The recent experiments have demonstrated the active control of triplet supercurrents in different types of S/F/S junctions\cite{Birge1, Birge2, Birge3, Birge4, Birge5} including the Josepshon systems with half-metallic layers.\cite{Keizer, Anwar} which provides the guidance for the appropriate choice of materials and their parameters.

To demonstrate the controllable switching of vorticity and the magnetic flux pumping we consider the superconducting loop of inductance $L$ interrupted by the Josephson $\psi$ junction characterized by the the current-phase relation $I=I_c\sin(\varphi-\psi)$ where $I_c$ is the critical current (see Fig.~\ref{Fig1}). The dynamics of the Josephson phase $\varphi$ is described by the sine-Gordon equation \cite{Barone}:
\begin{equation}\label{SineGordon}
\frac{\partial^2\varphi}{\partial t^2}+\frac{1}{R_NC}\frac{\partial\varphi}{\partial t}+\frac{1}{LC}\varphi+\frac{2\pi I_c}{\Phi_0 C}\sin\left[\varphi-\psi(t)\right]=0,
\end{equation}  
where $C$ and $R_N$ are the capacity and normal state resistance of the Josephson junction, respectively. Considering the case of identical easy-axis anisotropy of magnetization in both F layers of the $\psi$ junction we assume that in the absence of external driving $\psi(t)=0$. To analyze the system dynamics it is convenient to introduce the dimensionless time $\tau=\omega_p t$ where $\omega_p=\sqrt{2\pi I_c/\Phi_0 C}$ is the plasma frequency, the quality factor $Q=R\sqrt{2\pi I_c C/\Phi_0}$ and the normalized loop inductance $\lambda=2\pi LI_c/\Phi_0$.

\begin{figure}[t!]
\includegraphics[width=0.5\textwidth]{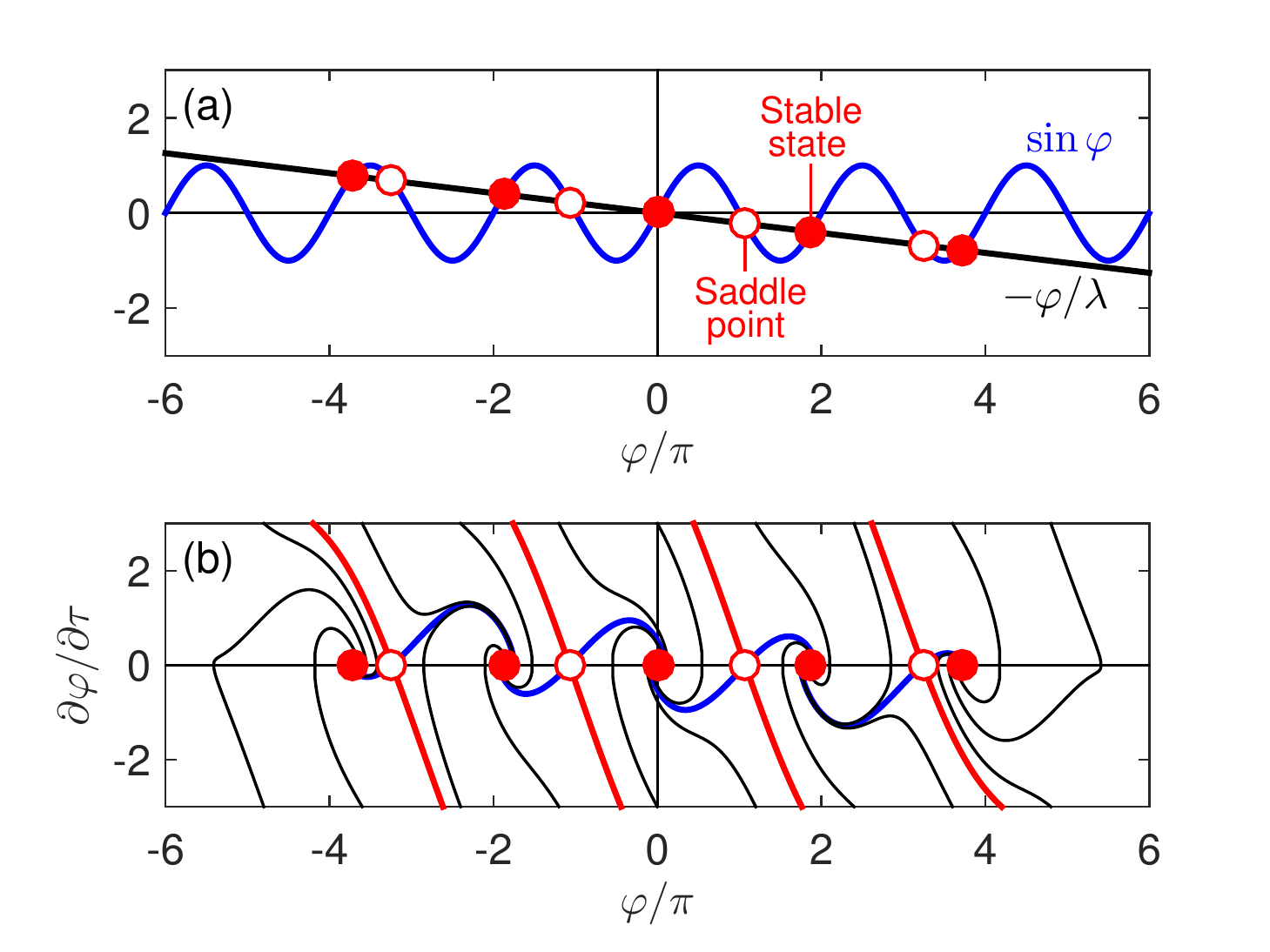}
\includegraphics[width=0.5\textwidth]{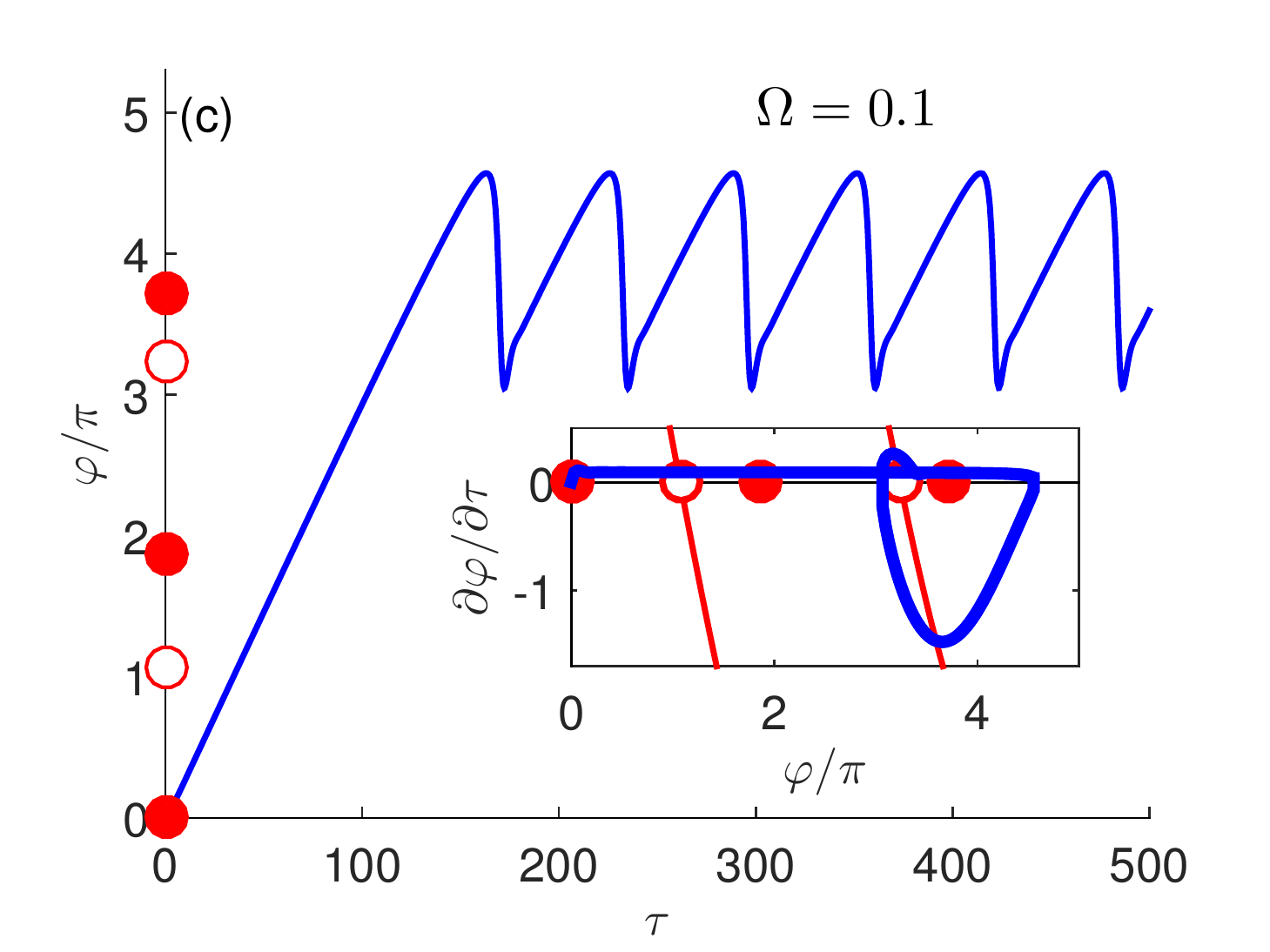}
\caption{(a) Stable states (filled red circles) of the system separated by the saddle points (empty red circles). (b) Phase portrait of the system. Red lines are the separatrices of the saddle points. (c) Dependence $\varphi(\tau)$ and the phase portrait (see inset) under the driving at $\Omega=0.1$ in the flux pumping regime. In all panels $\lambda = 15$ and $Q=1$.}\label{Fig3}
\end{figure}

In the absence of external driving the number of equilibrium superconducting states in the loop is controlled by the loop inductance. At small $\lambda$ the equation $\sin\varphi=-\varphi/\lambda$ defining the equilibrium states has only one trivial solution $\varphi=0$. When increasing $\lambda$ at the threshold values $\lambda=\lambda_n$ one gets the sequential emergence of additional pairs of the stable states $\varphi=\pm \varphi_n$ ($n=1,2,3...$) and the nearby saddle points $\varphi=\pm\varphi^{sdl}_{n}$ (see Fig.~\ref{Fig3}a, b). The precession of the magnetic moment with the frequency $\omega$ in one of the F layers produces the subsequent growth of the Josephson ground-state phase $\psi=\omega t$ and the dynamics of $\varphi$ is controlled by the equation
\begin{equation}\label{SGdim}
\frac{\partial^2\varphi}{\partial \tau^2}+\frac{1}{Q}\frac{\partial\varphi}{\partial \tau}+\frac{1}{\lambda}\varphi+\sin(\varphi-\Omega\tau)=0,
\end{equation}  
where $\Omega=\omega/\omega_p$. Further we analyze the dynamics of the Josephson phase $\varphi$ numerically solving Eq.~(\ref{SGdim}) with the explicit Euler method.

The periodic driving with rather small $\Omega$ being applied to the system at the initial equilibrium state $\varphi=0$ produces almost linear growth of $\varphi$ accompanied by the increase in the magnetic flux through the loop (see Fig.~\ref{Fig3}c). The described flux pumping is limited by the maximal stable Josephson phase, and the further precession of the magnetic moment gives rise to the periodic change of $\varphi$ around the maximal stable state $\varphi_n$.

\begin{figure}[hbt!]
\includegraphics[width=0.5\textwidth]{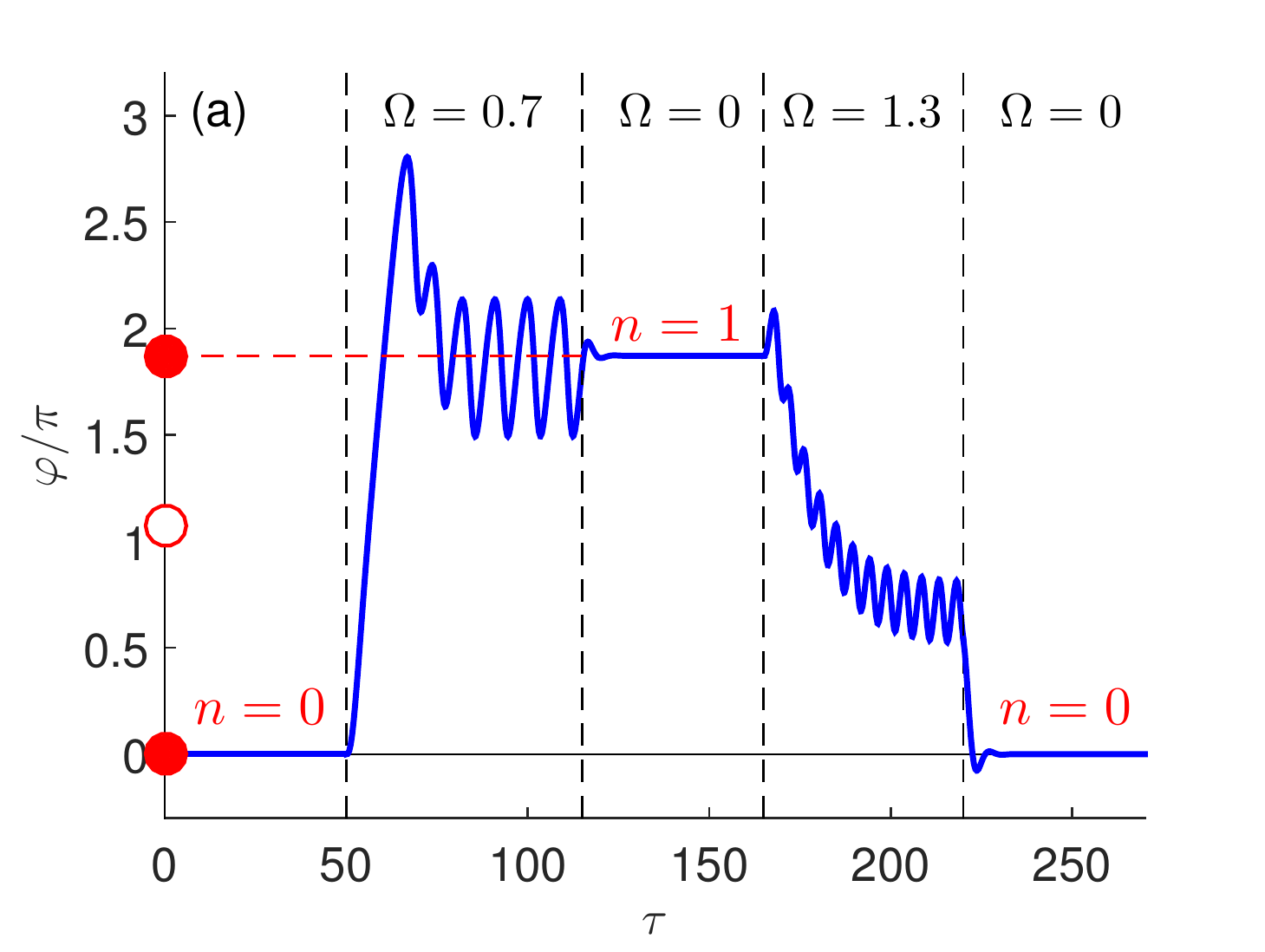}
\includegraphics[width=0.23\textwidth]{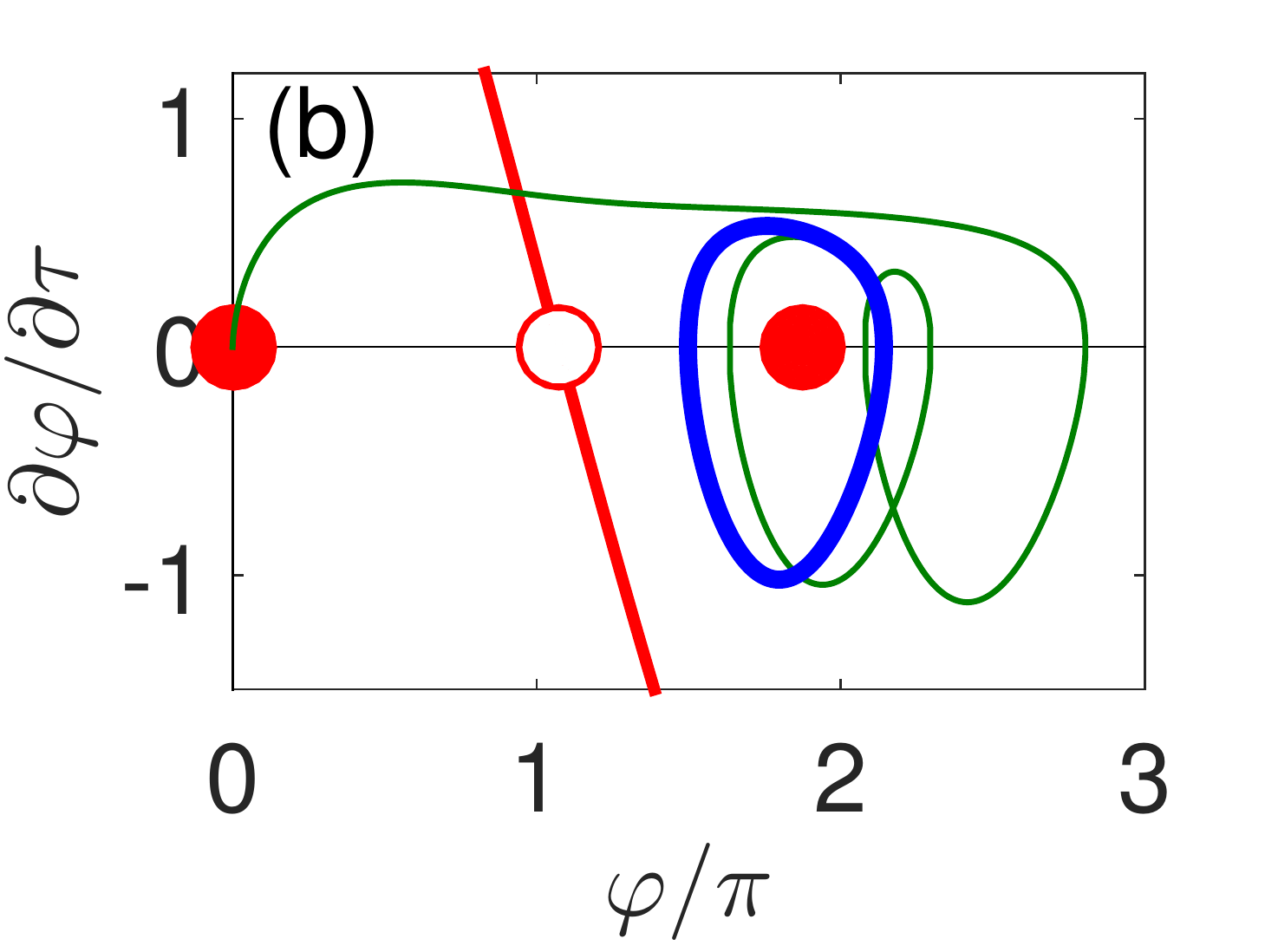}
\includegraphics[width=0.23\textwidth]{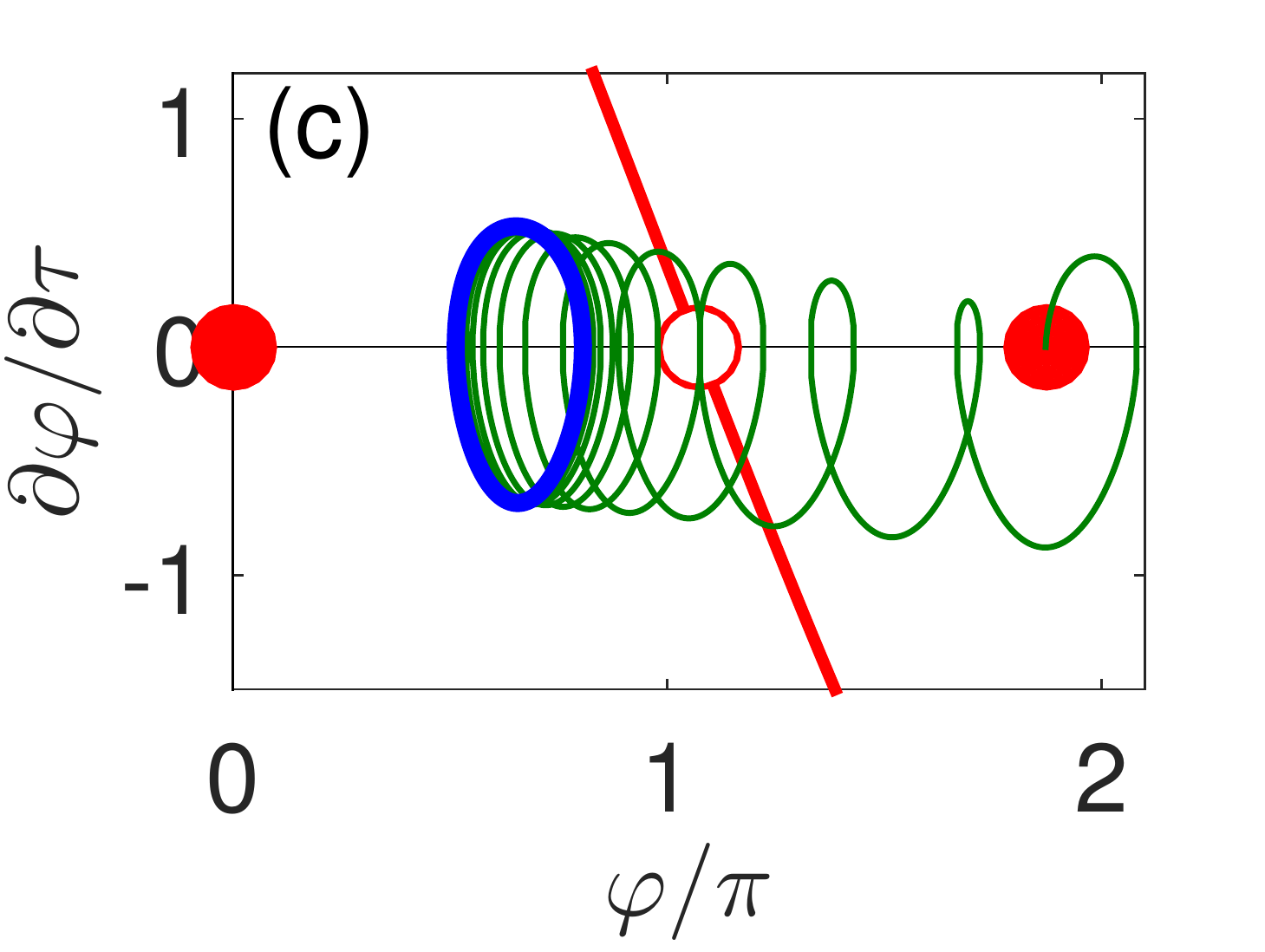}
\caption{(a) Dynamics of the Josephson phase corresponding to the controllable switching from the state $n=0$ to the state $n=1$ and back. The applied driving frequencies are indicated on top of the figure. The dashed lines indicate the moments when the driving is switched on and off. (b), (c) Corresponding phase portraits of the system under the driving at $\Omega=0.7$ and $\Omega=1.3$, respectively. We took $\lambda=15$ and $Q=1$.} \label{Fig4}
\end{figure}

The appropriate choice of the driving frequency $\Omega$ allows the controllable switching between different stable states $\varphi_n$. Let us assume that initially the system is in the stable state $\varphi=0$ and the periodic driving is switched on at $\tau=0$. The solution of Eq.~(\ref{SGdim}) with the initial conditions $\varphi(0)=0$ and $\partial_\tau\varphi(0)=0$ at large $\tau$ relaxes to the periodic oscillations of Josephson phase and, thus, the oscillations of the magnetic flux through the loop (see Fig.~\ref{Fig4}a). The corresponding phase portrait on the plane $(\varphi,\partial_\tau \varphi)$ has the form of the closed cycle (see Fig.~\ref{Fig4}b) which position at the phase diagram is controlled by the driving frequency $\Omega$. For the appropriate choice of $\Omega$ the phase trajectory crosses one of separatrices and the whole limiting cycle lays between two neighboring separatrices. In this case switching off the driving at any arbitrary moment with 100\% probability should result in the further relaxation of the system to the state which is different from the initial one. Note that the external driving with another frequency $\Omega$ can return the system back to the initial state, so that the switching between the states with different $n$ is reversible. The described scenario is illustrated in Fig.~\ref{Fig4} where the periodic driving with the frequency $\Omega=0.7$ switches the system from the state with $n=0$ to the state with $n=1$ while the further driving with the frequency $\Omega=1.3$ produces the inverse switching. 

Note that if the limiting cycle crosses one of the separatrices the final state strongly depends on the specific moment when the driving is switched off which makes such regime unfavorable for the design of the switching devices. 

\begin{figure}[t!]
\includegraphics[width=0.5\textwidth]{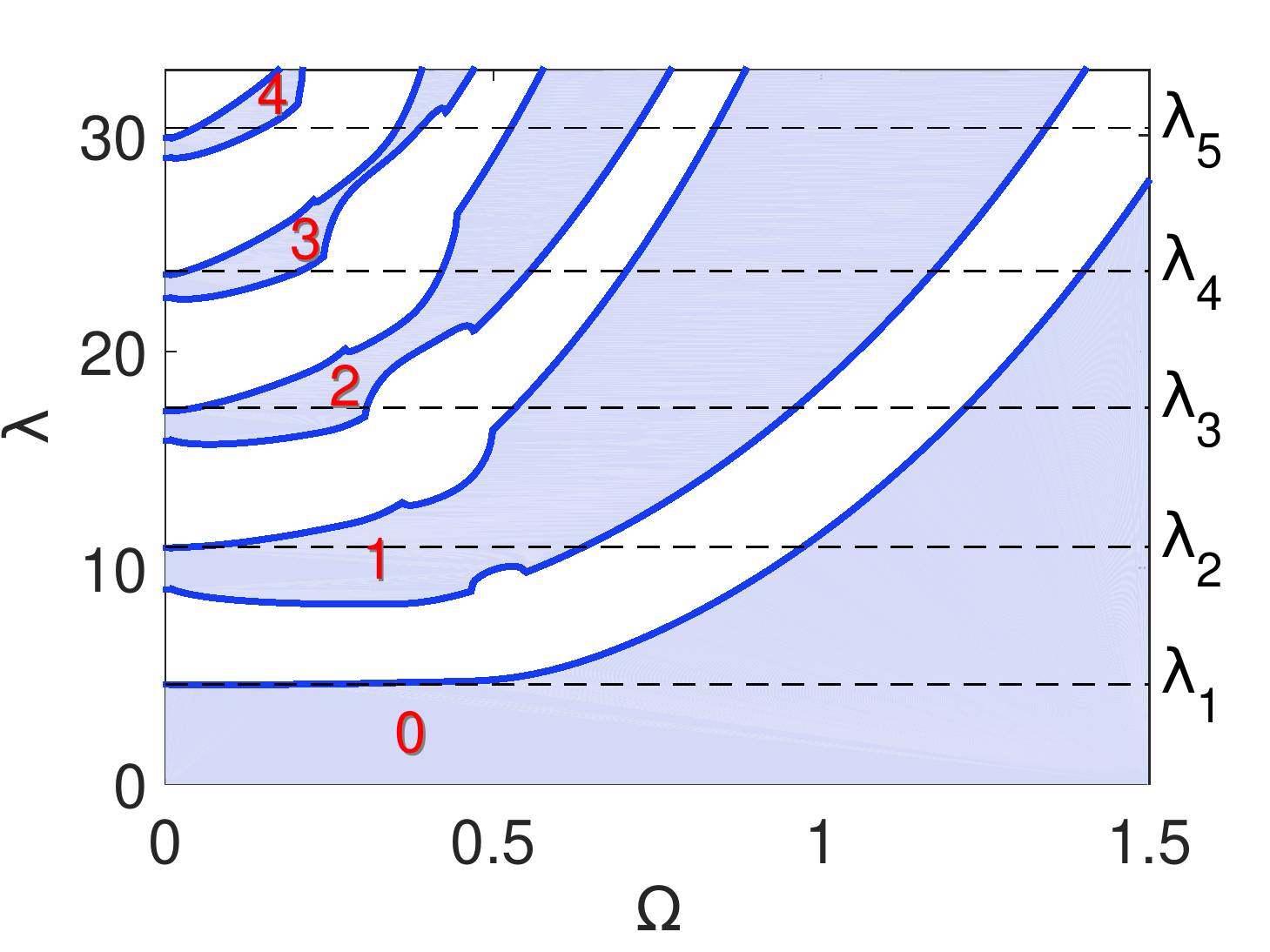}
\caption{Diagram of the switching regimes controlled by the driving frequency $\Omega$. Each point inside the filled blue areas guarantees the switching to the stable state $\varphi_n$ with the fixed vorticity (the corresponding index $n$ is shown with the red color). For the parameters in the intermediate white regions the final state depends on the specific moment when the driving is switched off. The values $\lambda_n$ corresponding to the appearance of additional stable states are shown with the horizontal black lines. } \label{Fig5}
\end{figure}

In Fig.~\ref{Fig5} we provide the diagram guiding the choice of the driving frequency $\Omega$ corresponding to the 100\% probability switching. If for the fixed normalized loop inductance $\lambda$ the driving frequency is chosen inside the filled blue region then the limiting cycle does not cross the separatrices and after the driving is switched off the system would relax to the state $\varphi_n$ (the corresponding index $n$ is written inside the filled blue region). In contrast, the choice of $\Omega$ in the intermediate white region would produce the cycle crossing the separatrices and, thus, cannot guarantee the controllable switching of the state vorticity. 

Note that throughout the paper we focused on the periodic variations of the Josephson phase stimulated by linearly growing $\psi(t)$. At the same time, for $\omega\lesssim\omega_p$ with the increase of the quality factor $Q$ the periodic dependence $\varphi(t)$ first experiences the series of the period doubling bifurcations and then becomes non-periodic which is rather typical for the rf SQUIDs.\cite{Huberman, Kautz} Exactly such doubling of the period is responsible for the curvature inversion of the region boundaries in Fig.~\ref{Fig5}. However, to distinguish whether the non-periodic behavior of $\varphi(t)$ is truly chaotic or not the further investigation is needed.

We have considered the influence of the magnetic moment re-orientation on the Josephson phase. Note that the inverse effect, when the voltage applied to the $\psi$ junction produces the precession of the magnetic moment, is also possible. In such a case we may expect the magnetic moment re-orientation, similar to the situation considered in Refs.~\onlinecite{Shukrinov_1,Shukrinov_2} for the $\varphi_0$ junction.

Finally, let us estimate the parameters of the superconducting loop and the Josephson $\psi$ junction required for the realization of the described memory cell.  For the typical Josephson junctions in the RSFQ devices with the critical current $I_c\sim 1~ {\mu A}$, the normal state resistance $R_N\sim 100~{\rm\Omega}$ and the capacitance $C\sim 1~{\rm pF}$ the plasma frequency lays in the ${\rm GHz}$ range ($\omega_p\sim 10^{10}\div 10^{11}~{\rm sec}^{-1}$) and the quality factor $Q\sim 1$. Then from Fig.~\ref{Fig5} one sees that the existence of at least one stable state with $n=1$ requires $\lambda\sim 10$ which corresponds to the loop inductance $L\gtrsim 10^{-11}~{\rm H}$ (rather typical values for the existing devices). The operating frequencies $\omega$ then should be of the order of $\omega_p$. Comparing required $\omega$ values with the frequencies $\omega_f\sim 10^{11}~{\rm sec}^{-1}$ of the magnetization precession induces, e.g., by the circularly polarized radiation \cite{Kimel_Nat} one sees that the proposed mechanism of the vorticity switching can be realized in the currently fabricated RSFQ circuits.

Obviously, the effects of the finite temperature and noise which were not taken into account should modify the systems dynamics. However, the estimates\cite{Likharev} show that the effects of temperature-induced fluctuations do not result in qualitative changes provided the loop inductance L is small compared to the threshold value $L_c\sim(\Phi_0/2π)^2 (1/k_B T)\sim 10^{-9}~{\rm H}$ at $T = 4~{\rm K}$. Thus, for rather small loops of the inductance $L\gtrsim 10^{-11}~{\rm H}$ the effects of fluctuation should not be crucial.

To sum up, we have demonstrated that the integration of the Josephson S/F/HM/F/S $\psi$ junction into the superconducting loop enables the controllable switching between the superconducting states with different vorticities without application of external magnetic flux. The peculiar coupling between the  precession of the magnetic moment in the F layer and the Josephson phase oscillations in the $\psi$ junction allow the realization of the magnetic flux pumping into the loop controlled, e.g., by the microwave radiation. The proposed mechanism may open the avenue for the next generation of the memory cells and the operating devices of the RSFQ based electronics.

\acknowledgments 

The authors thank V. V. Ryazanov, A. S. Mel'nikov and A. A. Fraerman for useful discussions. This work was supported by  the French ANR SUPERTRONICS and OPTOFLUXONICS, EU COST CA16218 Nanocohybri, and the Russian Foundation for Basic Research (grant 18-02-00390). In part related to the diagram of the controlled flux switching the work was supported by the Russian Science Foundation (grant 18-72-10027). S. M. acknowledges the financial support of the  Foundation for the advancement of theoretical physics “BASIS” and Russian Presidential Scholarship SP-3938.2018.5.  H.M. acknowledges the financial support from National Natural Science Foundation of China (grant 11604195), and Youth Hundred Talents Program of Shaanxi Province.

The data that support the findings of this study are available from the corresponding author upon reasonable request.

\end{document}